\documentclass[%
 reprint,
 amsmath,amssymb,
 aps,
]{revtex4-2}

\usepackage{graphicx}
\usepackage{dcolumn}
\usepackage{bm}
\usepackage{amsmath}
\usepackage{subcaption}

\begin{document}

\preprint{APS/123-QED}

\title{Mutual information scrambling in Ising spin chain}

\author{Surbhi Khetrapal$^{a,b}$ }
\email{surbhikhetrapal@uohyd.ac.in}
 \author{Emil Tore Mærsk Pedersen$^{b,c}$}
 
 \affiliation{$^a$School of Physics, University of Hyderabad, Gachibowli 500046 Telangana,\\
 $^b$Vrije Universiteit Brussel, Brussels 1050 Belgium,\\
 $^c$Department of Mathematical Sciences, University of Copenhagen, 2100 Copenhagen, Denmark}

\begin{abstract}
We consider a chain of spin-$1/2$ particles of a finite length $L$ evolved with the mixed-field Ising Hamiltonian and impose open boundary condition. We simulate the time evolution of entanglement entropy and mutual information following quench from the N\'eel state in this system using tensor networks. We find that the entanglement entropy for non-integrable systems saturates to a constant value at late times, however it continues to oscillate for integrable systems. We also find that mutual information peaks as a function of distance between intervals decay faster for non-integrable systems compared to integrable systems, in agreement with the conclusion of \cite{Alba:2019ybw} for XXZ chains. We compare the oscillations in entanglement entropy evolution obtained from simluations in the integrable case with analytic results from quasi-particle picture and find agreement.
\end{abstract}

\maketitle

\section{\label{sec1} Introduction}

Entanglement has become an increasingly important way of characterizing quantum many-body systems. An important question is how entanglement can be created and spreads starting from a system in a non-entangled product-state. Recent results have shown that the time-evolution of entanglement following quench can be an interesting tool to distinguish integrable and chaotic systems \cite{Kim_2013,Alba:2019ybw,Alba:2020yyh}. One way to quantify entanglement is via the entanglement entropy and mutual information.
The Ising model is a simple model of a one-dimensional spin-1/2 chain with only nearest-neighbor interactions, which can show both integrable and chaotic behaviour depending on values of the free parameters in the Hamiltonian. 

Investigating the scrambling of quantum information (also known as chaos) in many-body quantum systems \cite{Xu:2018dfp, David:2017eno}, provides valuable insights into the post-quench evolution of a system. While various methods are available to understand the chaotic dynamics of these quantum systems, such as spectral statistics \cite{Craps:2019pma} and out-of-time ordered correlators \cite{Xu:2018xfz, Khetrapal:2022dzy}, recently scrambling of mutual information following quench has also emerged as a diagnostic of chaotic dynamics \cite{Alba:2019ybw}. In this work we study the scrambling of mutual information following quench for a finite size transverse field Ising spin chain with open boundary condition. The latter boundary condition implies that the quasi-particles are reflected from the spin chain boundary and this leads to interesting patterns such as oscillatory behaviour of the entanglement entropy and mutual information at late times in the integrable case. We find that mutual information as a function of distance between entanglement intervals decays faster in non-integrable systems compared to integrable ones. These results are in agreement with those obtained in \cite{Alba:2019ybw} for XXZ spin chain.

\subsection{\label{sec:level2} Setup}

In this work, we consider a chain of spin-$1/2$ particles, of length $L$ evolved with the mixed-field Ising Hamiltonian, 
\begin{align}
    H = \sum_{i=1}^L g \sigma_x^{[i]} + \sum_{i=1}^L h \sigma_z^{[i]}+\sum_{i=1}^{L-1} J \sigma_z^{[i]} \sigma_z^{[i+1]},
\end{align}
where $\sigma_x$, $\sigma_y$ and $\sigma_z$ are the Pauli matrices, $J$ is the strength of the interaction between neighboring spins, $h$ is the strength of the longitudinal magnetic field and $g$ is the strength of the transverse magnetic field. Following \cite{Kim_2013}, we consider the following, slightly modified, Hamiltonian
\begin{align} \label{eq:Ising_Hamiltonian}
    H =  \sum_{i=1}^L g \sigma_x^{[i]} + \sum_{i=1}^{L-1} h \sigma_z^{[i]} &+ (h-J)(\sigma_z^{[1]}+\sigma_z^{[L]}) \nonumber \\
    & + \sum_{i=1}^{L-1} J \sigma_z^{[i]} \sigma_z^{[i+1]},
\end{align}
where the longitudinal fields at the ends of the chain are reduced by $J$ to keep the ends more similar to rest of the chain and avoid edge-effects.
Depending on the values of $h$ and $g$ the Ising model can be integrable or non-integrable. With either of the fields off $(h = 0$ or $ g = 0)$, the model is integrable and analytic results for the entanglement entropy evolution can be obtained. Examples of non-integrable values we  use are
\begin{align}
    (h,g,J) & = \left( \frac{\sqrt{5}+1}{4}, \frac{\sqrt{5}+5}{8} ,1 \right), 
\end{align}
which is used in \cite{Kim_2013}, and
\begin{align}
    (h,g,J) = (0.5,-1.05,1),
\end{align}
which is highly chaotic \cite{Banuls_2011}.

A common way to bring a quantum system out of equilibrium is a quantum quench: when a quantum system is prepared in an eigenstate of a Hamiltonian $H_0$ and then evolved in time under a different Hamiltonian $H_1$, this is called a quantum quench. It is of significant interest in quantum information research to study the creation and spreading of entanglement following a quench from a non-entangled product state. In this paper we quench the system from an initial state where every spin is pointing in the opposite direction (up or down) from its neighbors, called the N\'eel state,
\begin{align}
  |\Psi_N\rangle = |0\rangle \otimes |1\rangle \otimes |0\rangle \otimes |1\rangle \otimes \dots.
\end{align}
The N\'eel state is an eigenstate of the Ising Hamiltonian with the transverse field $g = 0$ turned off. Hence, evolving this initial state with the mixed field Ising Hamiltonian with non-zero transverse and longitudinal fields ($g \neq 0, h \neq 0$) is an example of a quench.

\subsection{Entanglement entropy and mutual information}
For a pure state $|\Psi \rangle \in H_A \otimes H_B$ the density matrix is simply $\rho = |\Psi \rangle \langle \Psi|$. The bipartite von-Neumann entanglement entropy of subsystem $A$ is defined as,
\begin{align} \label{eq:EE}
    S(A) = - \text{Tr} \left( \rho_A \log \rho_A \right)
\end{align}
where $\rho_A = \text{Tr}_B (\rho)$ is the reduced density matrix of subsystem $A$. Performing the trace in \eqref{eq:EE} in the orthonormal basis of $ H_A$ where $\rho_A$ is diagonal, with eigenvalues $\lambda_k$, gives
\begin{align} \label{eq:EE_diag}
    S(A) = - \sum_k \lambda_k \log \lambda_k .
\end{align}
It can be proven that $\rho_B$ has the same eigenvalues as $\rho_A$ so $S(A) = S(B)$. The entanglement entropy $S(A) = S(B)$ is a measure of the amount of entanglement between subsystems $A$ and $B$. 

A useful quantity that can be defined in terms of the entanglement entropy is the mutual information. The mutual information of two subsystems $A$ and $B$ is defined as
\begin{align}
    I_{A:B} = S_A+S_B-S_{A \cup B}.
\end{align}
The mutual information is a measure of the amount of information subsystems $A$ and $B$ have in common (due to their entanglement).

\subsection{Quasi-particle picture}

Quasi-particles are excitations in many-body systems, that can propagate through the system behaving like particles. Integrable quantum many-body systems admit stable quasi-particles with infinite lifetime, while non-integrable quantum many-body systems only admit quasi-particles with finite lifetime. A quantum quench can act as a source of quasi-particles. When the initial state is a product state, only pairs of quasi-particles created at the same point are entangled. When these entangled particles move in opposite directions through the system they will then spread the entanglement. Considering a subsystem $A$, a pair of entangled quasi-particles will then contribute to the entanglement entropy of $A$ only when one of the particles is in $A$ and the other one outside. 

\begin{figure}
\includegraphics[scale = 0.3]{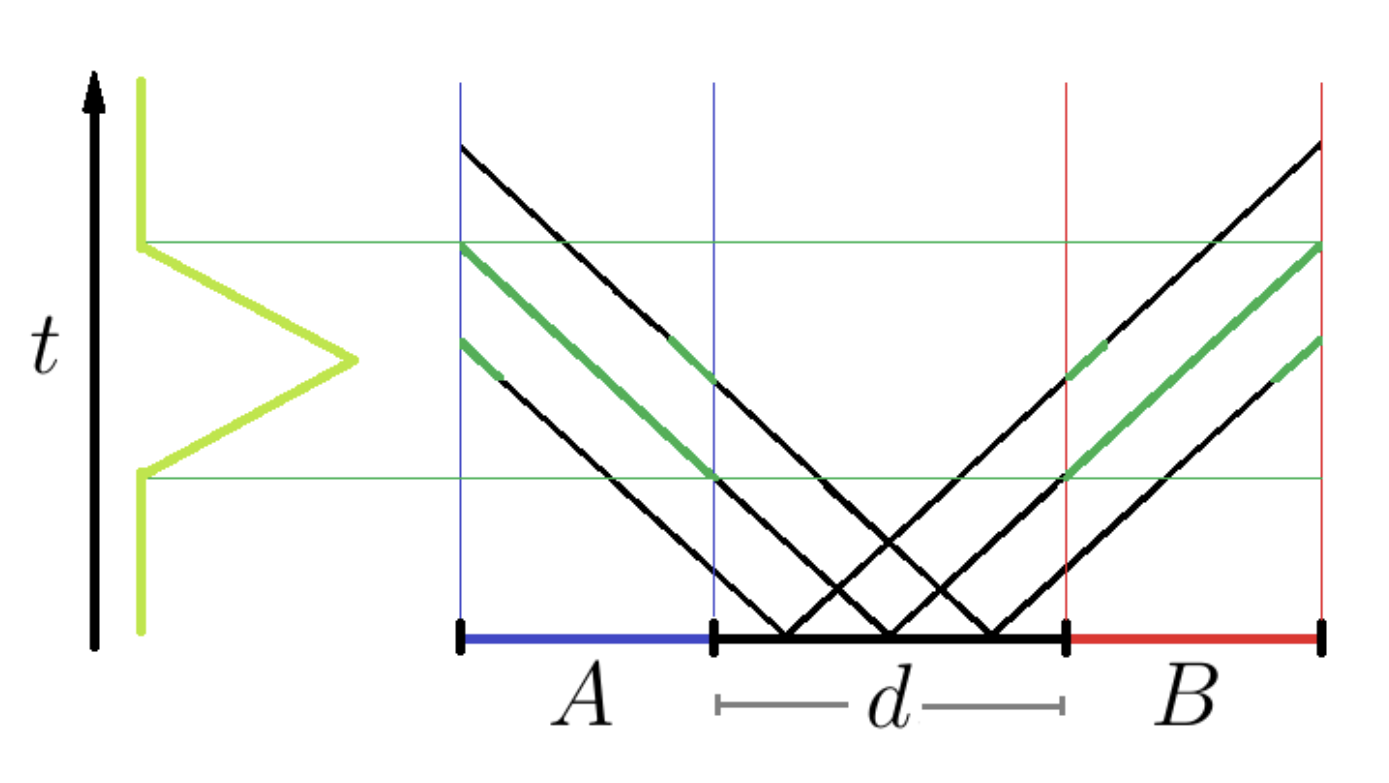}
\caption{\label{fig:evol_of_quasi} Quasiparticle picture of mutual information $I(A,B)$ time evolution}
\end{figure}

In case of the mutual information $I(A,B)$, a pair of quasi-particles will therefore contribute to the mutual information when one of the particles is in $A$ and the other in $B$. Figure \ref{fig:evol_of_quasi} shows a space-time diagram of a simple one-dimensional system where entangled pairs of quasi-particles (diagonal black world lines) moving in opposite directions, all with the same speed, are produced throughout the system from a quench at time $t = 0$. The three pairs shown in the figure, all reach a region in time (green part of their trajectory) where one of them is in $A$ and the other in $B$ and they therefore contribute to $I(A,B)$. This will result in a peak in $I(A, B)$ (light-green line in the figure, assuming pairs were created uniformly throughout the system) when a maximal amount of pairs contribute. In the more realistic case of a nontrivial distribution of quasi-particle species with different velocities, the mutual information will be a superposition of peaks from the different species.

\subsection{\label{sec:level2} Implementation using Tensor Networks}
Tensor network diagrams reviewed in \cite{ Schollw_ck_2011, Orus:2013kga,Orus:2018dya,Ran_2020} are a very convenient and elegant way to compute entanglement in quantum spin chains \cite{Vidal_2003,Vidal:2006ofj,vidal2009entanglement,Eisert:2013gpa,Banuls:2019qrq}. A program was written in Python for working with canonical Matrix Product States (MPS), implementing Time Evolving Block Decimation (TEBD) and computing the entanglement entropy for different subsystems. A few important considerations while implementing the algortihms described in these review articles are discussed below:
\begin{enumerate}
    \item A maximal bond-dimension D or cutoff value $\lambda_c= 10^{-16}$ was set for Schmidt coefficients $\lambda_k $ such that all Schmidt vectors and coefficients with $k \geq D$ or $\lambda_k < \lambda_c$, respectively were discarded efficiently with the NumPy arrays.

    \item Most of the results presented in this work are computed without limit on the bond dimension $D$. For a spin-$\frac{1}{2}$ chain, where each local site has a local state space of dimension $d = 2$, with even length $L$, the worst-case bond-dimensions required for an exact representation are
\begin{align}
    D_r = \begin{cases}
2^r & \text{when} \; r \leq \frac{L}{2}\\
2^{L-r} & \text{when} \; r > \frac{L}{2}
    \end{cases}.
\end{align}
Thus the maximum of these bond dimensions is $ D =  2^{\frac{L}{2}}$. For short chains this is not a problem on a fast processor but for longer chains, such as $L = 20$ or more, computational resources become a bottleneck, and efficient truncation is an important advantage of MPS. However, since our program was able to simulate these large lengths without truncation, we continued using exact bond-dimensions for the rest of the results.

    \item Computing the entanglement entropy for a bipartition of the form $([1 : r], [r+ 1 : N])$ is trivial, however for the more complicated case of computing the entanglement entropy of two spatially separated intervals $S_{A\cup B}$ (reduced density matrix $\rho_{A\cup B}$) with $A = [1 : a]$ and $B = [b : N]$, it is essential to use an efficient method to contract the central tensor, which can easily use far too much time or memory. We worked out and encoded the most efficient contraction order.

    \item In order to implement TEBD, it is necessary to bring the Ising Hamiltonian into a $2-$site nearest-neighbor form. Since the Hamiltonian is time-independent, these $h^{[r:r+1]}$ and their exponentials $U^{[r:r+1]} = e^{-\frac{i}{\hbar} \delta t h^{[r:r+1]}}$ can be computed from the parameters and the time-step $\delta t$ at the beginning of the computation and can then be reused in all TEBD steps ($\hbar$ is set to $1$). Following this, all time-evolution gates are applied in parallel at each even or odd step of TEBD. It is crucial to perform the tensor contractions efficiently and in the right order to minimize memory and time-usage.
\end{enumerate}

\section{\label{sec2} Simulation results}

\subsection{Evolution of Entanglement entropy after quench from N\'eel state}
We simulated chains of all even lengths from $L = 4$ to $20$ with the time step, $\delta t =0.01$ for several values of $h$ and $g$ to compare the results. The results for time evolution of the entanglement entropy of half-chain $S[1:L/2]$ following quench from N\'eel state and evolution by the Ising Hamiltonian \eqref{eq:Ising_Hamiltonian} are obtained.

\subsubsection{Integrable case}
\begin{figure}[htbp]
    \centering
    \begin{subfigure}[b]{0.45\textwidth}
        \centering
        \includegraphics[scale=0.57]{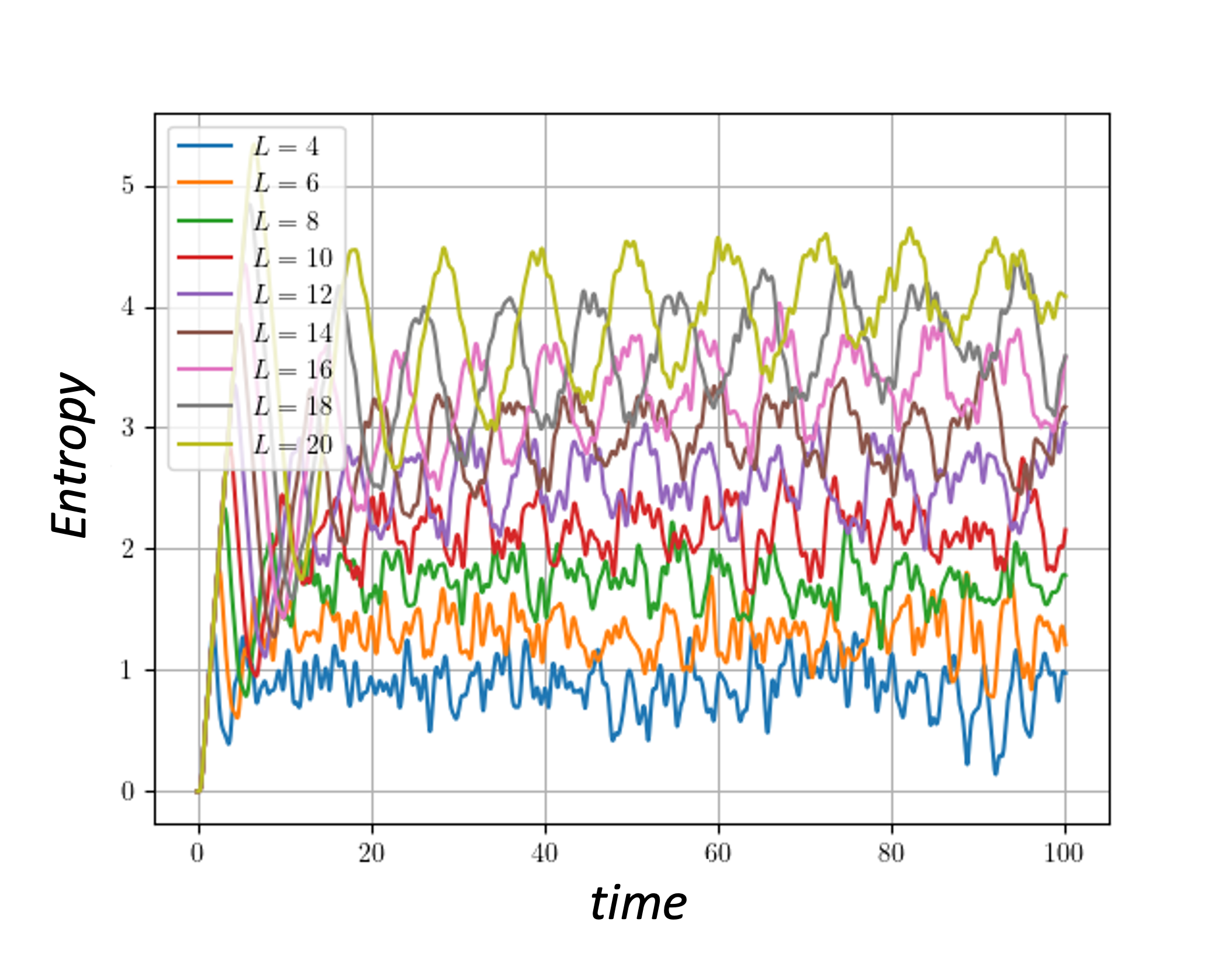}
        \caption{ \label{Fig:EE_int}  Integrable case $(h,g,J) = (0,1,1)$.}
    \end{subfigure}
    \hfill
    \begin{subfigure}[b]{0.45\textwidth}
        \centering
        \includegraphics[scale=0.57]{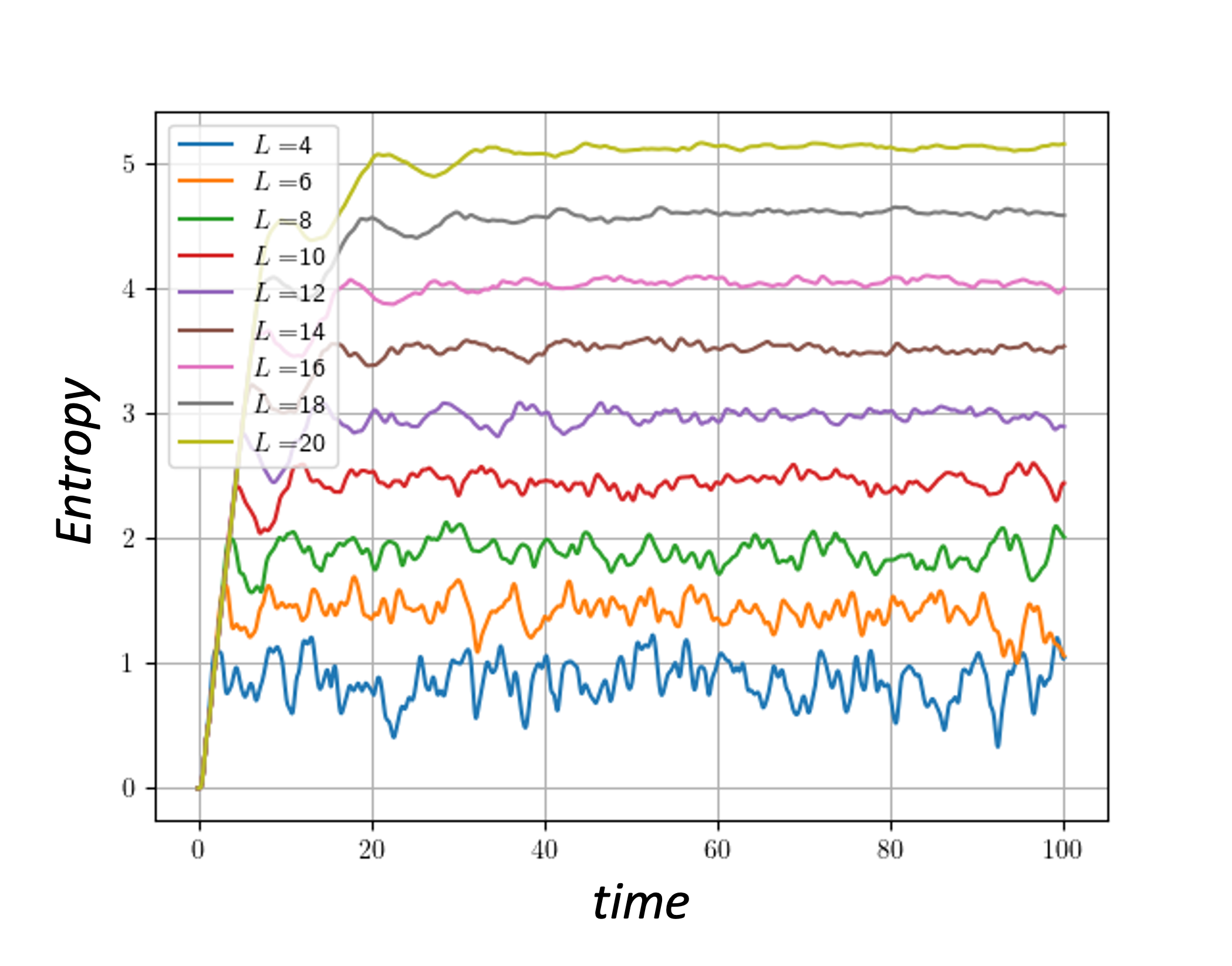}
        \caption{\label{Fig:EE_nonint} Non-integrable case $(h,g,J) = (\frac{\sqrt{5}+1}{4},\frac{\sqrt{5}+5}{8},1)$. }
    \end{subfigure}
    \caption{Time evolution of the entanglement entropy of half-chain $S([1:L/2])$ in spin chains with lengths $L=4$ to $20$.}
\end{figure}

As an example of an integrable case we considered the case with the longitudinal field off and transverse field on $(h,g,J) = (0,1,1)$ (for $g = 0$ the N\'eel state is an eigen-state of the Hamiltonian). The results for all lengths of spin-chains $L = \left\lbrace 4,6,\dots 20\right\rbrace$ are shown in Figure \ref{Fig:EE_int}. The initial growth is very similar to the non-integrable cases, but afterwards the behaviour differs: the initial growth ends at a clear peak followed by a rapid (approximately linear) decrease and then rise to a second peak. These oscillations then continue, becoming more irregular over time. The oscillations remain regular for longer times for longer chains. These oscillations in the evolution of entanglement entropy are called revivals and are a signature of an integrable system \cite{Alba:2020yyh} and are not present in non-integrable or chaotic systems as we will see next. An analytic understanding of revivals in entanglement entropy evolution is provided in section \ref{sec3}.

\subsubsection{Non-integrable case}
 As examples of non-integrable cases, we considered the Hamiltonian \eqref{eq:Ising_Hamiltonian} with parameters $(h,g,J) =  \left( \frac{\sqrt{5}+1}{4}, \frac{\sqrt{5}+5}{8} ,1 \right)$ and $(h, g, J ) = (0.5, -1.05, 1)$. Figure \ref{Fig:EE_nonint} shows the results for the former case for spin-chain lengths  $L = \left\lbrace 4, 6,\dots 20 \right\rbrace$. These agree with the results obtained by exact diagonalisation for lengths $L = \left\lbrace 4, 6, 8, 10, 12 \right\rbrace$ \cite{Tori2019}. All lengths start with a similar rapid growth of entanglement which saturates near time $t_L \approx \frac{L}{2}$ and then continues fluctuating irregularly around the saturation value near $S(t \gg t_L) \approx \frac{L}{4}$ . These fluctuations are very pronounced for short lengths and decrease for larger lengths. Similar behaviour is obtained for $(h, g, J ) = (0.5, -1.05, 1)$ however the initial growth is slightly faster and saturation values are slightly higher.

\subsection{Evolution of mutual information after quench from Neel state}
Mutual information scrambling following quench in the Heisenberg XXZ chain was studied in \cite{Alba:2019ybw}, comparing the behaviour for integrable versus non-integrable parameters. They look at the mutual information $I(A,B)$ between two intervals $A = [1 : 2]$ and $B = [L-1 : L]$ at the ends of the chain, and consider the behaviour of the mutual information as a function of the distance $d = L-4$ between the intervals, arguing that this can be used to distinguish the integrable from non-integrable cases. We simulated the time-evolution of the mutual information for the same intervals $A = [1 : 2]$ and $B = [L-1 : L]$ for the Ising Hamiltonian \eqref{eq:Ising_Hamiltonian}, with both the integrable and non-integrable values of parameters, $(h,g,J)$, to see how their results compare to the Ising chain.

\begin{figure}[htbp]
    \centering
    \begin{subfigure}[b]{0.45\textwidth}
        \includegraphics[scale=0.55]{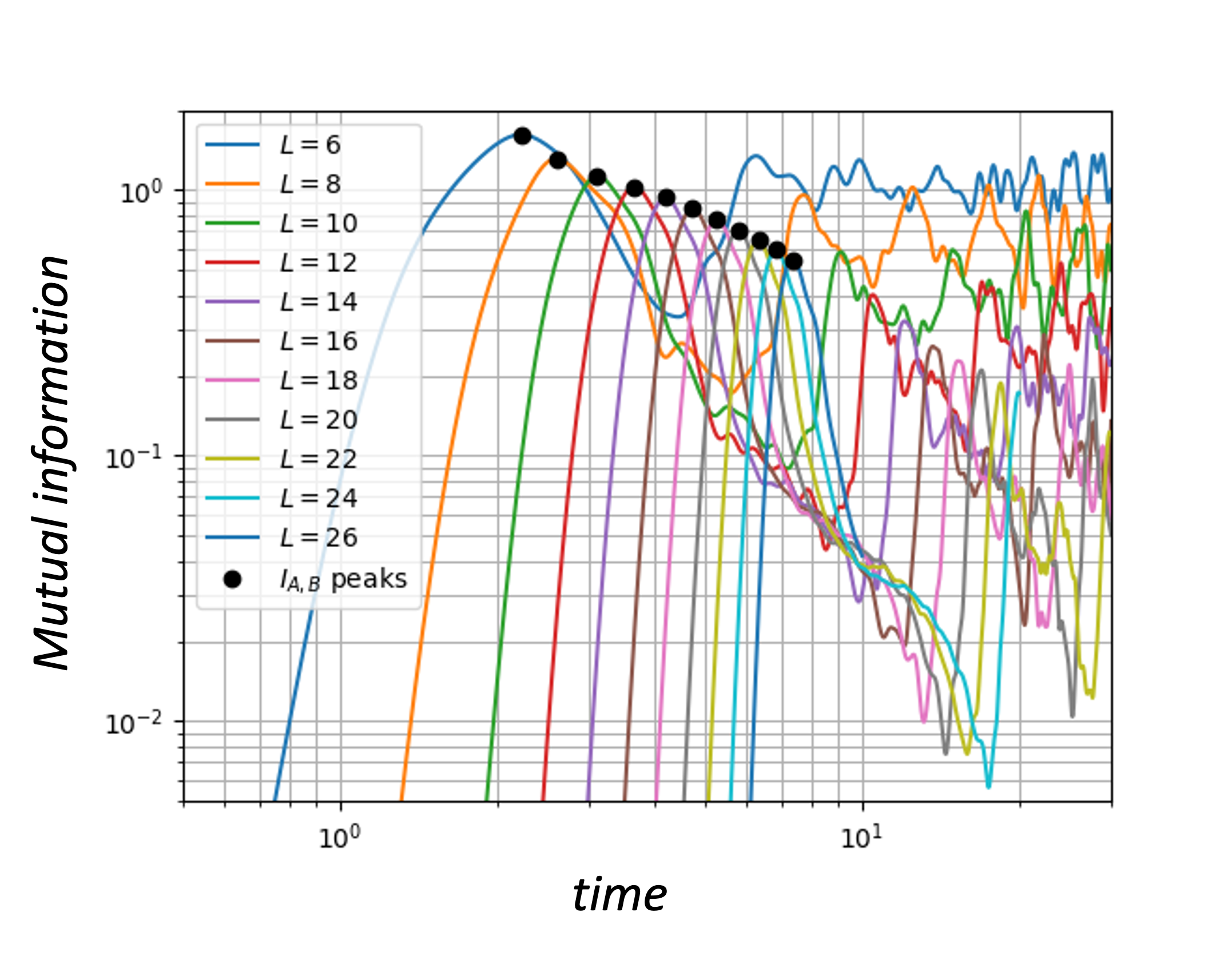}
        \caption{Integrable case $(h,g,J) = (0,1,1)$. }
        \label{Fig:MIpeaks_int}
    \end{subfigure}
    \hfill
    \begin{subfigure}[b]{0.45\textwidth}
        \includegraphics[scale=0.55]{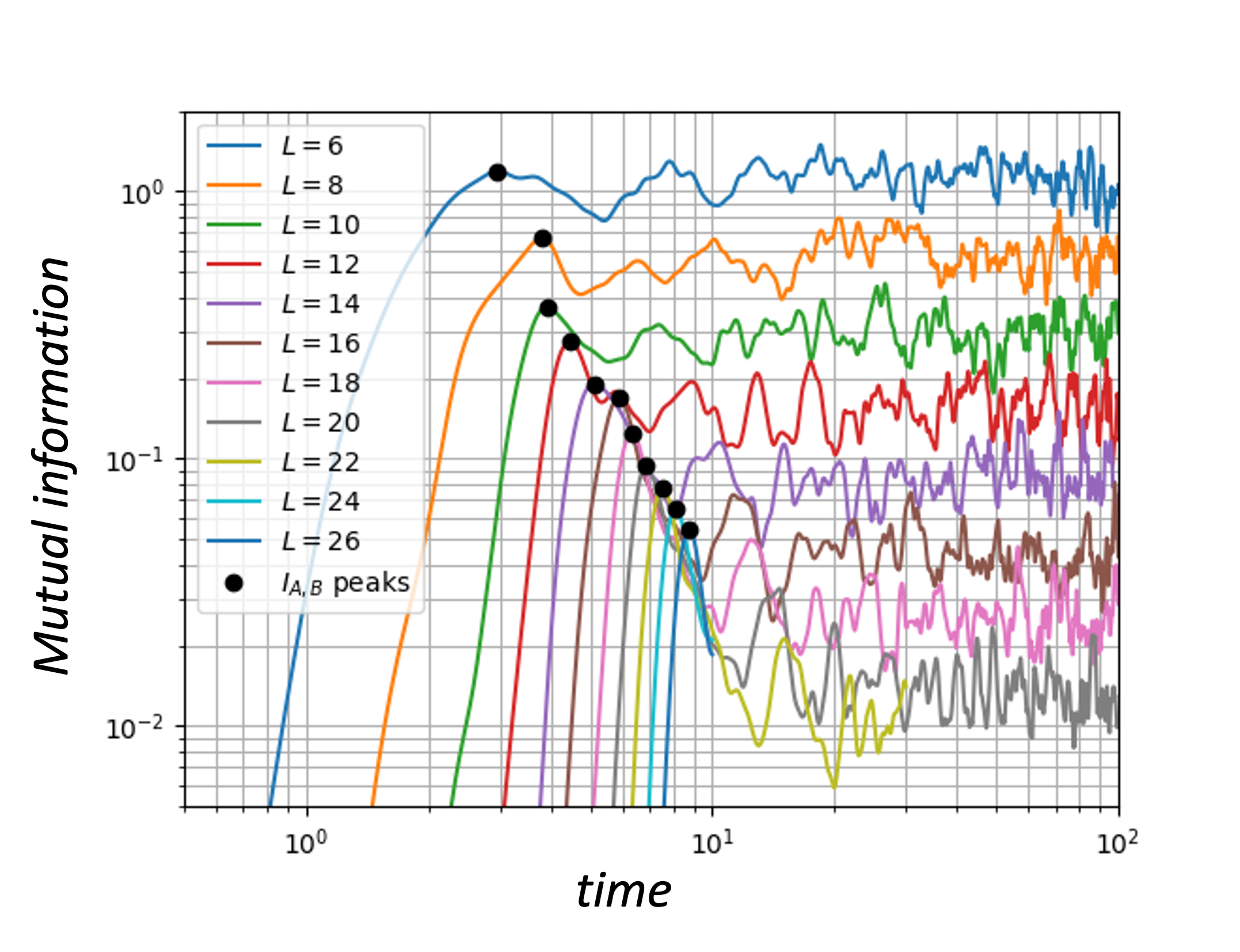}
        \caption{Non-integrable case $(h,g,J) = \left( \frac{\sqrt{5}+1}{4}, \frac{\sqrt{5}+5}{8} ,1 \right)$.}
        \label{Fig:MIpeaks_nonint}
    \end{subfigure}
    \caption{Time-evolution of mutual information, $I ([1 : 2], [L - 1 : L])$ in spin chains of lengths $L = 6$ to $26$. The above plots are in log-log scale for $t \leq 30$.}
\end{figure}

\begin{figure}
\centering
\includegraphics[scale = 0.5]{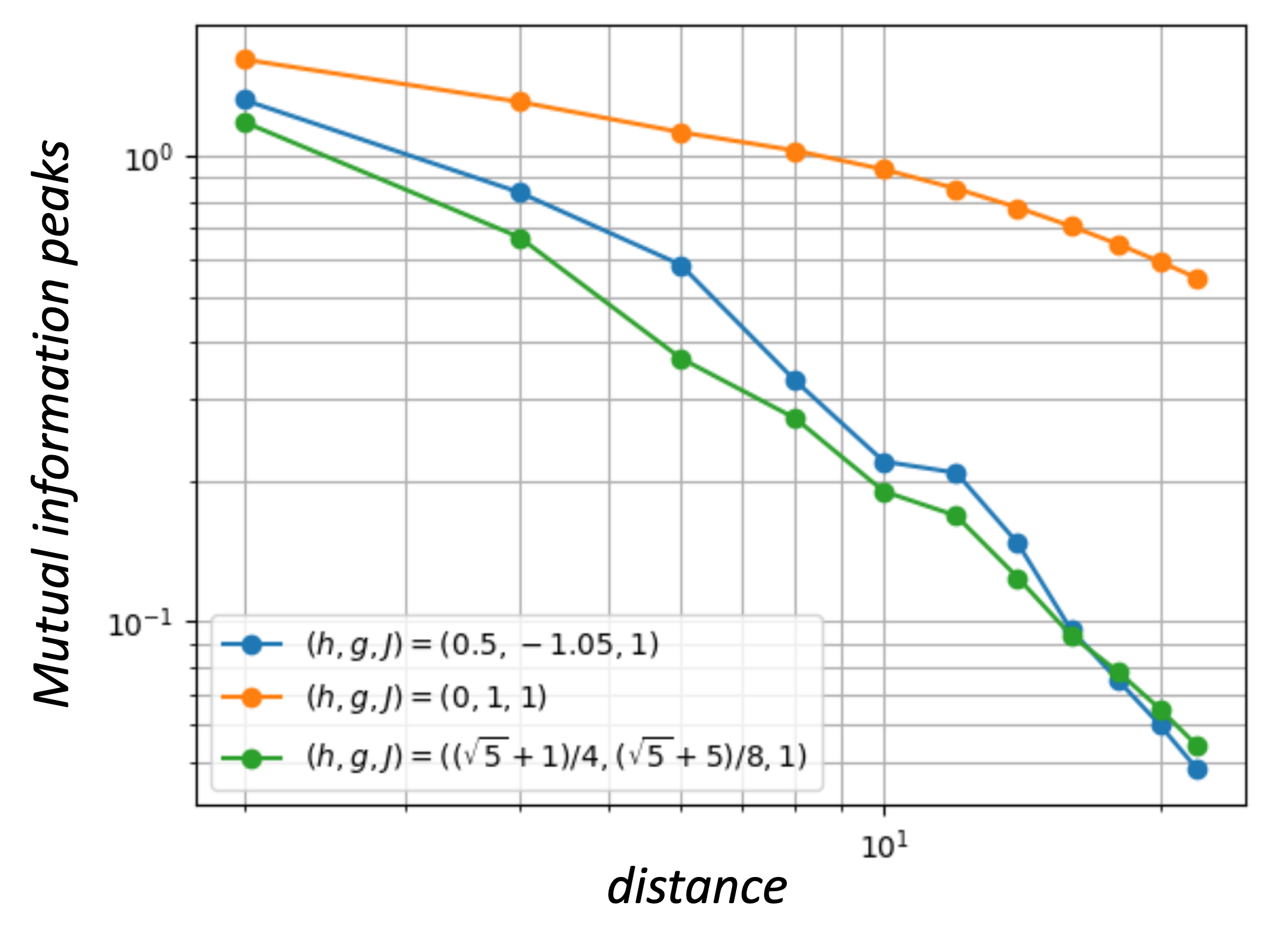}
\caption{\label{Fig:MIpeaks_all} Primary peak-value of mutual information $I([1 : 2],[L-1 : L])$ in function of distance $d = L - 4$, for different $(h, g, J)$. Here, the black dots mark the maxima of the first peak.}
\end{figure}

\subsubsection{Integrable case} \label{sec:results_int}
Figure \ref{Fig:MIpeaks_int} shows the results for the time-evolution of the mutual information $I ([1 : 2], [L - 1 : L])$, for lengths $L = 6$ to $26$ in the integrable case $(h,g,J) = (0,1,1)$. (Shorter chains were simulated until time $t = 100$, but length $L = 22$ was simulated until time $t = 50, L = 24$ until $ t = 40$, and $L = 26$ until $t = 10$). For all lengths, the mutual information first remains very low, followed by a rapid rise to a high peak (the black dots mark the maxima of the first peak in the figures). The time at which the peak occurs increases approximately linearly with $L$, while the height of the peak decreases with $L$. The presence of these peaks could be explained by the existence of a main species of quasi-particle with maximal velocity. For longer lengths $L$ the particles from the middle of the chain have to travel a longer distance $\approx \frac{L}{2} -1$ before they contribute to the mutual information, which means the peak will occur later in time and, if they are scrambled on the way, the height of the peaks will decrease. The first peak is followed by more peaks at regular intervals, with smaller peaks scattered in between. A striking feature of the main peaks is the regular intervals between them: for a fixed $L$, letting $t_0(L)$ be the time of the primary peak, there are always other large peaks at times $t_n(L) \approx (2n + 1)t_0(L), \; n \in \mathbb{N}$. This timing pattern is similar to the timings of the peaks in the evolution of the entanglement entropy $S([1 : L/2])$ (Figure \ref{Fig:EE_int}). 
Another interesting feature, mainly present in longer chains is that, after the big decrease from the primary peak to the next main peak, the height of the peaks temporarily rises with time. The regular timings of the main peaks could also be explained if the main species of quasi-particle, causing the primary peak, can elastically reflect off the edges of the chain and contribute to the mutual information again when they reach the opposite ends of the chain.

\subsubsection{Non-integrable case}
Figure \ref{Fig:MIpeaks_nonint} shows the results for the time-evolution of the mutual information $I([1 : 2], [L - 1 : L])$, for lengths $L = 6$ to $26$, with $(h,g,J) =  \left( \frac{\sqrt{5}+1}{4}, \frac{\sqrt{5}+5}{8} ,1 \right)$ and similar results were obtained for $(h,g,J) = (0.5,-1.05,1)$. (In both cases, shorter chains were simulated until time $t = 100$, but length $L = 22$ was simulated until time $t = 50$, $L = 24$ until $t = 40$, and $L = 26$ until $t = 10$). In both the cases, like in the integrable case, the mutual information first remains very low and then has a peak (black dots in the figures). Compared to the integrable case, the time at which the peak occurs is similarly linear in $L$, but the height of the peak decreases more rapidly with $L$. Figure \ref{Fig:MIpeaks_all} shows the heights of these peaks as a function of the distance $d = L - 4$ between the intervals, for both the non-integrable cases and the integrable case. The presence of these well-defined peaks suggests there might still be a main species of quasi-particle in the non-integrable cases, but the rapid decrease in height (compared to the integrable case) suggests they are scrambled significantly faster than in the integrable case. After the primary peak there are no other clear regular peaks like in the integrable case.

The behaviour of the peaks as a function of $d$ (Figure \ref{Fig:MIpeaks_all}) and the difference between the non-integrable and integrable cases agrees well with the results of \cite{Alba:2019ybw}, where they similarly observed faster decrease in the non-integrable cases, and slower (algebraic $\propto d^{-1/2}$) decrease for the integrable cases. Therefore they argued that non-integrable systems are better scramblers than integrable systems, and that scrambling is an interesting way of distinguishing integrable and chaotic systems. \cite{Alba:2019ybw} didn’t find many examples of peaks other than the primary ones, this could be due to shorter simulation times than our results.

\section{\label{sec3} Entanglement entropy from quasi-particle picture for a finite system}

We briefly review the computation of entanglement entropy from quasi-particle picture developed in \citep{Calabrese:2005in}. The initial state $|\psi_0\rangle$ has very high energy relative to the ground state of the Hamiltonian, $H(\lambda)$. Therefore, it acts as a source for quasi-particle excitations. These particles from non-coincident points do not entangle, but particles moving to the left or right from a given point are highly entangled. The cross-section to produce pairs of particles of momenta $(k',k'')$ is $f(k',k'')$. Once these quasi-particles separate they move classically. The quasi-particle velocity is $v(k) = \frac{d\epsilon_k}{dk}$, where $\epsilon_k$ is its dispersion relation. Thus the maximum allowed speed is $|v(k)| \leq 1$.

A quasi-particle with momentum $k$ produced at $x$ is at $x+v(k)t$ at a given time $t$. As quasi-particles reach either $A$ or $A^c$ at time $t$, they are entangled if the particles reaching simultaneously at $x' \in A$ and $x'' \in A^c$ is a pair emitted from $x$. Thus the entanglement entropy is proportional to the length of the interval in $x$ for which this can be satisfied,

\begin{widetext}
\begin{align}
S_A(t) = \int_{x' \in A} dx' \int_{x'' \in B} dx'' \int_{-\infty}^{\infty} dx \int f(k',k'') dk' dk'' \delta \left(x'-x-v(k')t \right) \delta \left( x''-x-v(k'')t\right).
\end{align}
\end{widetext}

Here $A$ is an interval of length $l$. The total entanglement is twice that between $A$ and the part of $A^c$ to the right of $A$. Thus restricting to the later implies that $k'<0, k''>0$. Performing the integral over $x,x'$ and $x''$ gives, $\text{max}\left( t \left(v(-k')+v(k'') \right),l \right)$. Substituting this, the entanglement entropy becomes,

\begin{widetext}
\begin{align}
S_A(t) = &  \int_{-\infty}^0 dk' \int_0^\infty dk'' f(k',k'') \left(v(-k')+v(k'') \right) \left[  2t \, \theta \left(l-\left(v(-k')+v(k'') \right)t \right)  + 2l \, \theta \left(\left(v(-k')+v(k'') \right)t-l \right) \right].
\end{align}
\end{widetext}
In \cite{Alba:2017lvc}, it was proposed that for a periodic system, using $k''=k, k'=-k$, the entanglement entropy is,
\begin{align}
S_A(t) = 2t \int_{2|v|t <l} dk v(k)s(k)+l \int_{2|v|t>l} dk s(k)
\end{align}
in the limit that $t,l \to \infty$ with $r/t$ fixed.

Next, the entanglement entropy from the quasi-particle picture is derived for a large system size of length $L$, where $t,l,L \to \infty$ such that $t/l$ and $l/L$ (or $t/L$) are kept fixed. The entanglement entropy for periodic boundary condition is \cite{Modak:2020faf},
\begin{align}\label{eq:EE_periodic}
S_l(t) = & \int_{ \left\lbrace \frac{2v(k)t}{L} \right\rbrace<\frac{l}{L}} \frac{dk}{2\pi} s(k) L \left\lbrace \frac{2v(k)t}{L} \right\rbrace \nonumber\\ 
& + l \int_{ \frac{l}{L} \leq \left\lbrace \frac{2v(k)t}{L} \right\rbrace < 1-\frac{l}{L}} \frac{dk}{2\pi} s(k) \nonumber\\
& +  \int_{ 1-\frac{l}{L} \leq \left\lbrace \frac{2v(k)t}{L} \right\rbrace}  \frac{dk}{2\pi}s(k) L \left( 1-\left\lbrace \frac{2v(k)t}{L} \right\rbrace \right)
\end{align}
where $\{ x \}$ means the fractional part of $x$.

The entropy grows linearly up to $\frac{l}{2v}$, then exhibits a plateau which terminates at $t=\frac{(L-l)}{2v}$. The plateau terminates when the first quasi-particle produced at one boundary of the sub-system re-enters it from another edge after having turned around the circle. Subsequently, more and more quasi-particles re-enter the system causing the entropy to decrease till the revival time $t_R = \frac{L}{2v}$. At the revival time, the dynamics begins again as if the system were at $t=0$. Further computations to obtain entanglement entropy by performing the above integrals are given in appendix \ref{app1}.

The results of simulations presented in section \ref{sec:results_int} are for open boundary condition. The above computation can be modified to give the entanglement entropy in the case of open boundary condition by rescaling both $L \to 2L$ and $l \to 2l$. This is because in the case of open boundary condition, the quasi-particles reflect from the boundary of the spin chain and the dynamics is the same as if the system size is doubled and periodic boundary condition are imposed. Thus the revival time in this case is $t_R = \frac{L}{v}$. 
\begin{figure}
    \centering
    \includegraphics[scale = 0.5]{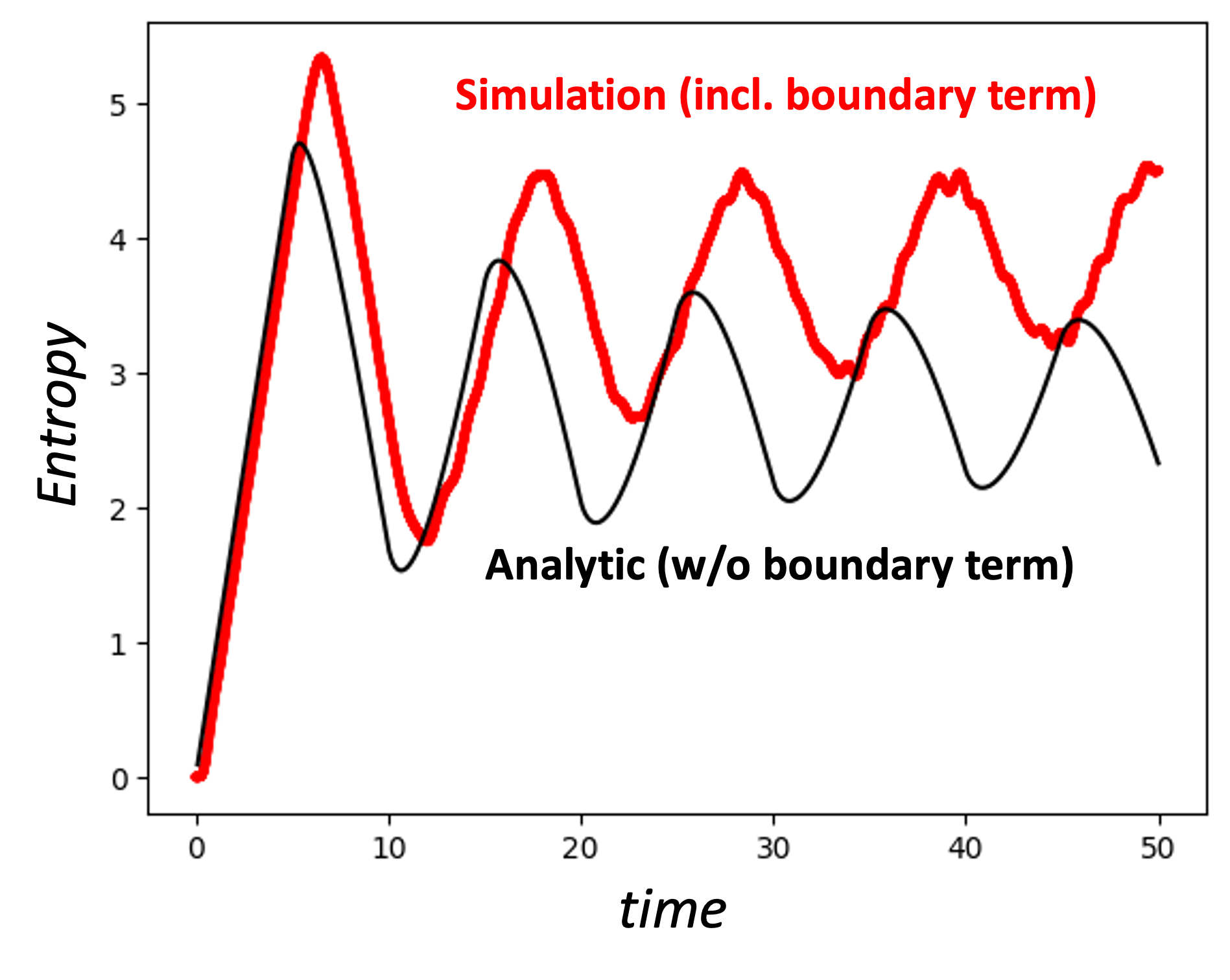}
    \caption{Revivals in entanglement entropy following quench for the integrable Ising hamiltonian with $(h,g,J) = (0,1,1)$. The black line depicts the analytic result and the red line depicts the result of simulation. The length of the spin chain is $L=20$ and the entanglement entropy is computed of half-chain of length $l=10$ and $v_m = 2$. The difference in the results is due to boundary effects.}
    \label{Fig:EE_integrable_analytic}
\end{figure}
Figure \ref{Fig:EE_integrable_analytic} shows revivals in entanglement entropy following quench for the integrable Ising hamiltonian with open boundary condition for $(h,g,J) = (0,1,1)$. The black line depicts the analytic result and the red one depicts the result of simulation. The length of the spin chain is $L=20$ and the entanglement entropy is computed of the half-chain of length $l=10$. The difference in the results is due to boundary effects. In the analytic case, the reduction in magnitude of the longitudinal field at the boundary points has not been taken into account. Also, there may be slowing down of quasi-particles due to reflection at the boundaries of this open spin chain, this effect is captured in the entanglement entropy computed by simulation (as opposed to analytic computation) and hence gives a higher value at the first peak in entanglement entropy and a longer revival time.

These results as well as the ones in \cite{Alba:2020yyh} (for finite integrable spin chains) differ from \cite{Calabrese:2005in} where the authors consider an infinite integrable system and thus obtain a linear rise followed by saturation of entanglement entropy as a function of time. Thus these revivals in evolution of entanglement entropy for the integrable spin chains is a finite-size effect.

\section{\label{sec4} Outlook}
In this paper, we saw that, mutual information as a function of distance between the entanglement intervals decays faster in chaotic spin chains compared to integrable ones. It is known that the dynamics of an Ising model at the critical point may be descibed by a certain conformal field theory \cite{Iqbal:2020msy, Zamolodchikov:1989mz}. In a conformal field theory, in order to compute the R\'enyi mutual information of two spatially separated intervals, we compute the $n$-th R\'enyi entropies, which are obtained from correlation functions of twist operators,
\begin{align}
    I^{(n)}_{A:B} = S^{(n)}_A+S^{(n)}_B-S^{(n)}_{A \cup B}
\end{align}
where
\begin{align}
 e^{(1-n) S^{(n)}_A} & =   \left\langle \sigma_n \left(-\frac{a}{2} \right){\bar{\sigma}}_n\left(\frac{a}{2} \right)\right\rangle \\
 e^{(1-n)S^{(n)}_B} & =  \left\langle \sigma_n \left(d-\frac{a}{2} \right){\bar{\sigma}}_n\left(d+\frac{a}{2} \right)\right\rangle \nonumber \\
 e^{(1-n)S^{(n)}_{A \cup B}} & =  \left\langle \sigma_n \left(-\frac{a}{2} \right){\bar{\sigma}}_n\left(\frac{a}{2} \right) \sigma_n \left(d-\frac{a}{2} \right){\bar{\sigma}}_n\left(d+\frac{a}{2} \right) \right\rangle.\nonumber
\end{align}
 Hence,
\begin{align}
    e^{(n-1) I^{(n)}_{A:B}} =\frac{\left\langle \sigma_n \left(-\frac{a}{2} \right){\bar{\sigma}}_n\left(\frac{a}{2} \right) \sigma_n \left(d-\frac{a}{2} \right){\bar{\sigma}}_n\left(d+\frac{a}{2} \right) \right\rangle}{\left\langle \sigma_n \left(-\frac{a}{2} \right){\bar{\sigma}}_n\left(\frac{a}{2} \right)\right\rangle \left\langle \sigma_n \left(d-\frac{a}{2} \right){\bar{\sigma}}_n\left(d+\frac{a}{2} \right)\right\rangle}
\end{align}
Thus the R\'enyi mutual information of two intervals is the four-point function of twist operators in the orbifold CFT \cite{Chen_2017, Calabrese_2009, Calabrese_2011}. This correlator can be written as a sum over conformal blocks \cite{Chen_2017,Rajabpour:2011pt},
\begin{align}
    & \left\langle \sigma_n \left(-\frac{a}{2} \right){\bar{\sigma}}_n\left(\frac{a}{2} \right) \sigma_n \left(d-\frac{a}{2} \right){\bar{\sigma}}_n\left(d+\frac{a}{2} \right) \right\rangle \nonumber \\
    & = c_n^2 x^{-\frac{c}{6}(n-\frac{1}{n})} \left( \sum_K \alpha_K d^2_k x^{h_K} F(h_K,h_K;2h_K;x)\right)^2,
\end{align}
where $x$ is the cross-ratio, $F(h_K,h_K;2h_K;x)$ is the hypergeometric function, summation over $K$ is over the independent quasi-primary operators and other constants are to be determined from the operator product expansion of the twist fields.

When using the above formula to compute the four-point function of twist operators, one can only compute sum over a certain finite number of blocks, thus this method provides an approximation to the correlator. While the correlator is single valued, the sum over blocks is multi-valued, this leads to several possible channels via which the four twist operators may be contracted to obtain an approximation to the above correlator. One of these channels will bring the twist operators in out-of-time ordering in imaginary time. In this channel we expect the mutual information to be an exponentially decaying function of the distance between entanglement intervals, $d$. In addition, similar to the bound on Lyapunov exponent in holographic theories \cite{Maldacena:2015waa}, we expect that the rate of exponential decay of mutual information as a function of the distance between entanglement intervals, $d$, will be bounded by it's value in Einstein gravity. It will be interesting to find this decay rate and compare it with the one obtained for chaotic spin chains in this paper.

\section*{Acknowledgments}
This work began as part of the bachelor project of Emil Tore Mærsk Pedersen at Vrije Universiteit Brussel (VUB). Part of the results in this paper were obtained during his bachelor project and were presented in his (unpublished) bachelor project report \cite{emilthesis}, 2020. We thank Ben Craps and Charles Rabideau for discussions during the course of the bachelor project. We also thank Diptiman Sen, Nisheeta Desai and Tanay Nag for helpful comments and questions. Research of SK during postdoctoral tenure at VUB was supported by FWO-Vlaanderen
project G006918N and by Vrije Universiteit Brussel through the Strategic Research
Program High-Energy Physics. Currently, SK’s research is supported by Department
of Science and Technology’s INSPIRE grant DST/INSPIRE/04/2020/001063. ETMP's research is currently supported by Novo Nordisk Foundation (Grant No. NNF20OC0059939 ‘Quantum for Life’).

\onecolumngrid
\appendix

\vspace{0.5in}

\section{\label{app1}  Analytic results for entanglement entropy evolution in integrable systems}
\subsection{Entanglement entropy for $XX$ chain}
The XY chain is described by the Hamiltonian:
\begin{align}\label{eq:XY_Hamiltonian}
H_{XY} = - \sum_{j=1}^L \left[ \frac{1+\gamma}{2} S^x_j S^x_{j+1} +\frac{1-\gamma}{2} S^y_j S^y_{j+1} +h S^z_j \right]
\end{align}
with periodic boundary conditions, and it can be mapped to the free fermion model given by:
\begin{align}
H = \sum_k \epsilon_k c^\dagger_k c_k 
\end{align}
where $\epsilon_k$ represents the dispersion relation and is related to the velocity $v(k)$ as follows:
\begin{align}
\epsilon^2_k = (h-\cos k)^2 + \gamma^2 \sin^2 k, \nonumber\\
v(k) = \frac{d\epsilon_k}{dk}.
\end{align}
Upon a quench $h_0 \to h$ and $\gamma_0 \to \gamma$, the Bogoliubov angle is given by:
\begin{align}\label{eq:Bogoliubov_angle}
\cos \Delta_k = \frac{h h_0 -\cos k (h+h_0) + \cos^2 k+\gamma \gamma_0 \sin^2 k}{\epsilon \epsilon_0}.
\end{align}
Utilizing the Generalized Gibbs Ensemble (GGE), the entanglement entropy can be expressed as,
\begin{align}\label{eq:GGE_entropy}
s(k)= - \frac{1+\cos \Delta_k}{2} \ln \left( \frac{1+ \cos \Delta_k}{2}\right)- \frac{1-\cos \Delta_k}{2} \ln \left( \frac{1- \cos \Delta_k}{2}\right).
\end{align}

Now, let's examine a N\'eel quench in an $XX$ chain. The N\'eel state corresponds to $h_0=-\infty, \gamma_0=0$, while the $XX$ chain corresponds to $h=\gamma=0$. Consequently,
\begin{align}
\epsilon_0 = h_0 -\cos k, \qquad \epsilon=-\cos k,
\end{align}
and the velocity for both dispersion relations is $v(k) = v_M\sin k$, with $v_M=1$. The entropy $s(k)$ is constant at $s=\ln 2$, for all $k$.

\begin{figure}[h!]
    \centering
    \includegraphics[width=0.5\textwidth]{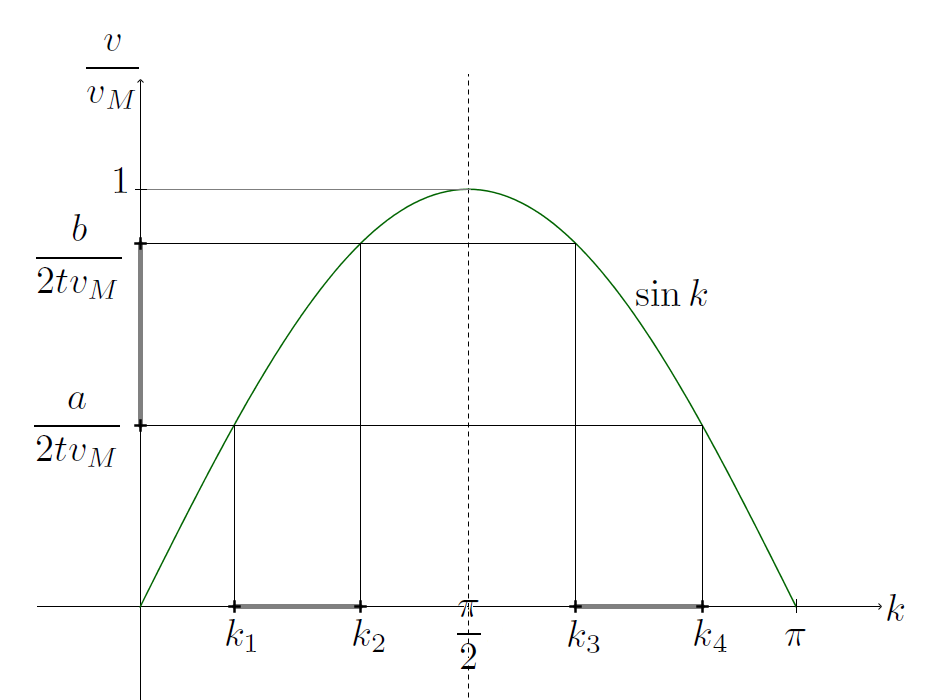}
    \caption{Integration boundaries $\frac{a}{2t} \leq v_M\sin k < \frac{b}{2t}$}\label{fig:XXintegralBC}
\end{figure}

First, considering the case when $t<t_R$, then $\left\lbrace \frac{2v(k)t}{L} \right\rbrace = \frac{2v(k)t}{L}$, so the integral boundaries in (\ref{eq:EE_periodic}) become,
\begin{align}\label{eq:EE_periodic_substXX}
S_l(t) =\int_{v_M\sin k<\frac{l}{2t}} \frac{dk}{2\pi} 2stv_M \sin k +& l \int_{ \frac{l}{2t} \leq v_M\sin k < \frac{L-l}{2t}} \frac{dk}{2\pi} s \nonumber\\
+&  \int_{ \frac{L-l}{2t} \leq v_M\sin k < \frac{L}{2t}}  \frac{dk}{2\pi} s \left( L- 2tv_M\sin k \right).
\end{align}
The integration boundaries are illustrated in Figure \ref{fig:XXintegralBC}, where $k_1=\begin{cases}\frac{\pi}{2}& \mbox{if } t< \frac{a}{2v_M}\\ \sin^{-1}\left(\frac{a}{2tv_M}\right) & \mbox{if } t\geq \frac{a}{2v_M}\end{cases}$, $k_2=\begin{cases}\frac{\pi}{2}& \mbox{if } t< \frac{b}{2v_M}\\ \sin^{-1}\left(\frac{b}{2tv_M}\right) & \mbox{if } t\geq \frac{b}{2v_M}\end{cases}$, $k_3=\pi-k_2$ and $k_4=\pi-k_1$ . 

Utilizing these boundaries and denoting $t_R=\frac{L}{2v_M}$ and $t_S=\frac{l}{2v_M}$, the resulting expression for $S_l(t<t_R)$ is,
\begin{align}
S_l(t<t_R)=\begin{cases}
\frac{2v_Ms}{\pi}t &\mbox{if } 0\leq t < t_S\\
\frac{2v_Ms}{\pi}t\left(1-\sqrt{1-\left(\frac{t_S}{t}\right)^2}\right) + \frac{ls}{\pi}\left(\frac{\pi}{2}-\sin^{-1}\left(\frac{t_S}{t}\right)\right) &\mbox{if } t_S\leq t < t_R-t_S\\
\begin{aligned} &\frac{2v_Ms}{\pi}t\left(1-\sqrt{1-\left(\frac{t_S}{t}\right)^2} - \sqrt{1-\left(\frac{t_R-t_S}{t}\right)^2} \right) \\ &~~+ \frac{ls}{\pi}\left(\sin^{-1}\left(\frac{t_R-t_S}{t}\right)-\sin^{-1}\left(\frac{t_S}{t}\right)\right) + \frac{Ls}{\pi}\left(\frac{\pi}{2}-\sin^{-1}\left(\frac{t_R-t_S}{t}\right)\right)\end{aligned} &\mbox{if } t_R-t_S \leq t < t_R.\end{cases}
\end{align}

When $t_R\leq t<2t_R$, the fractional part $\left\lbrace\frac{2v(k)t}{L}\right\rbrace$ takes on two distinct cases:
$$\left\lbrace\frac{2v(k)t}{L}\right\rbrace = \begin{cases}
\frac{2v(k)t}{L} & \text{if } v(k)<\frac{L}{2t} \\
\frac{2v(k)t}{L}-1 & \text{if } v(k)\geq\frac{L}{2t}
\end{cases}.
$$

Splitting the integrals accordingly:
\begin{align}
& \int_{\frac{a}{L} \leq \left\lbrace \frac{2v(k)t}{L} \right\rbrace < \frac{b}{L}} f\left(\left\lbrace \frac{2v(k)t}{L} \right\rbrace\right)dk \nonumber \\
&= \int_{(\frac{a}{L} \leq \frac{2v(k)t}{L} < \frac{b}{L}) \land (v(k)<\frac{L}{2t})} f\left(\frac{2v(k)t}{L}\right)dk + \int_{(\frac{a}{L} \leq \frac{2v(k)t}{L}-1 < \frac{b}{L}) \land (v(k)\geq\frac{L}{2t})} f\left(\frac{2v(k)t}{L}-1\right)dk \nonumber \\
&= \int_{\frac{a}{2t} \leq v_M\sin k < \frac{b}{2t}} f\left(\frac{2v(k)t}{L}\right)dk + \int_{\frac{a+L}{2t} \leq v_M\sin k < \frac{b+L}{2t}}  f\left(\frac{2v(k)t}{L}-1\right)dk,
\end{align}
where $0\leq a \leq b \leq L$. The entanglement entropy is now,
\begin{align}
&S_l\left(t\in[t_R,2t_R]\right)= \frac{2v_Ms}{\pi}t\left(1-\sqrt{1-\left(\frac{t_S}{t}\right)^2} + 2\sqrt{1-\left(\frac{t_R}{t}\right)^2} - \sqrt{1-\left(\frac{t_R-t_S}{t}\right)^2} \right) \nonumber\\
&+\frac{ls}{\pi}\left(\sin^{-1}\left(\frac{t_R-t_S}{t}\right)-\sin^{-1}\left(\frac{t_S}{t}\right)\right)+\frac{Ls}{\pi}\left(2\sin^{-1}\left(\frac{t_R}{t}\right)-\sin^{-1}\left(\frac{t_R-t_S}{t}\right)\right) \nonumber\\
&+\begin{cases}
-\frac{Ls}{\pi}\frac{\pi}{2} &\mbox{if } t_R\leq t < t_R+t_S\\
\begin{aligned} &-\frac{2v_Ms}{\pi}t \sqrt{1-\left(\frac{t_R+t_S}{t}\right)^2} - \frac{Ls}{\pi}\sin^{-1}\left(\frac{t_R+t_S}{t}\right) \\ &~~+ \frac{ls}{\pi}\left(\frac{\pi}{2}-\sin^{-1}\left(\frac{t_R+t_S}{t}\right)\right)\end{aligned} &\mbox{if } t_R+t_S\leq t < 2t_R-t_S\\
\begin{aligned} &-\frac{2v_Ms}{\pi}t\left(\sqrt{1-\left(\frac{t_R+t_S}{t}\right)^2} + \sqrt{1-\left(\frac{2t_R-t_S}{t}\right)^2} \right) \\ &~~- \frac{Ls}{\pi}\left(\sin^{-1}\left(\frac{t_R+t_S}{t}\right)+2\sin^{-1}\left(\frac{2t_R-t_S}{t}\right)-\pi\right) \\ &~~+ \frac{ls}{\pi}\left(\sin^{-1}\left(\frac{2t_R-t_S}{t}\right)-\sin^{-1}\left(\frac{t_R+t_S}{t}\right)\right)\end{aligned} &\mbox{if } 2t_R-t_S \leq t < 2t_R. \end{cases}
\end{align}
Further intervals $nt_R \leq t < (n+1)t_R$, $n\in\mathbb{N}$, can be calculated analogously, each time building upon the results of the previous intervals.

\subsection{Modifying the above calculation for quench to Ising chain}
The transverse field Ising model Hamiltonian is,
\begin{align}\label{eq:Ising_Hamiltonian_app}
H_I = -J\sum_{j=1}^L \left(\sigma_j^z\sigma_{j+1}^z + g\sigma_i^x\right).
\end{align}
This is the special case of the XY Hamiltonian $H_{XY}=-\frac{1}{2} \sum_{j=1}^L \left[ \frac{1+\gamma}{2} \sigma^x_j \sigma^x_{j+1} +\frac{1-\gamma}{2} \sigma^y_j \sigma^y_{j+1} +h \sigma^z_j \right]$ when $\gamma=1$ and $h=g$, scaled by a factor $2J$, rotated by $\frac{\pi}{2}$ around the $y$-axis (so $\sigma^x \mapsto -\sigma^z , \sigma^z \mapsto \sigma^x$). (For $J=1, g=1$, the results obtained are the same as the integrable case considered in the main text.) Therefore the dispersion relation (invariant under the rotation) is
\begin{align}
\epsilon_k=2J\sqrt{1+g^2-2g\cos k},
\end{align}
and the velocity for this dispersion relation is
\begin{align}
v(k)=2J\frac{g\sin k}{\sqrt{1+g^2-2g\cos k}}.
\end{align}

We now consider the N\'{e}el quench in the Ising model with $g=1$. The N\'{e}el state corresponds to $g_0=0$, which now corresponds to $h_0=0,\gamma_0=1$ in the (rotated) XY chain. So the Bogoliubov angle difference (\ref{eq:Bogoliubov_angle}) is $\cos\Delta_k = \left|\sin\left(\frac{k}{2}\right)\right|$ and therefore,
\begin{align}\label{eq:Ising_s(k)}
s(k) = -\frac{1+\left|\sin\left(\frac{k}{2}\right)\right|}{2}\ln\left(\frac{1+\left|\sin\left(\frac{k}{2}\right)\right|}{2}\right)-\frac{1-\left|\sin\left(\frac{k}{2}\right)\right|}{2}\ln\left(\frac{1-\left|\sin\left(\frac{k}{2}\right)\right|}{2}\right)
\end{align}
and, for $g=1$,
\begin{align}\label{eq:Ising_v(k)}
v(k)=v_M\cos\left(\frac{k}{2}\right)\operatorname{sgn}k \qquad,\quad v_M=2J.
\end{align}
In the integrals of (\ref{eq:EE_periodic}) the value of $k\in[0,\pi]$, so we can ignore the absolute values in $s(k)$ and the  $\operatorname{sgn}k$ in $v(k)$ in the following calculation.

For all $m\in\mathbb{N}$,
\begin{align}
\frac{mL}{2t}\leq v <\frac{(m+1)L}{2t} \implies \left\lbrace\frac{2vt}{L}\right\rbrace = \frac{2vt}{L}-m,
\end{align}
so if $\frac{mL}{2t}\leq v <\frac{(m+1)L}{2t}$ then
\begin{align}
\left( \frac{a}{L} \leq \left\lbrace\frac{2vt}{L}\right\rbrace < \frac{b}{L} \right) \iff \left( \frac{a+mL}{2t} \leq v < \frac{b+mL}{2t}\right).\nonumber
\end{align}
Therefore,
\begin{align}
\int_{\frac{a}{L} \leq \left\lbrace\frac{2vt}{L}\right\rbrace < \frac{b}{L}} f(k)dk= \sum_{m} \int_{\frac{a+mL}{2t} \leq v < \frac{b+mL}{2t}} f(k)dk,
\end{align}
when $0\leq a\leq b \leq L$. It is also easily seen that when $t\leq nt_R$, $n\in\mathbb{N}$, the only nonzero terms in this sum are those with $m<n$. So (\ref{eq:EE_periodic}) can be written
\begin{align}\label{eq:EE_periodic_deltasum}
S_l(t\in[nt_R,(n+1)t_R]) &= \sum_{m=0}^n (\Delta A_m + \Delta B_m + \Delta C_m) \nonumber\\
&\begin{aligned}
\Delta A_m &= \int_{\frac{mL}{2t} \leq v < \frac{l+mL}{2t}}\frac{dk}{2\pi}s(k)(2tv(k)-mL) \\
\Delta B_m &= \int_{\frac{l+mL}{2t} \leq v < \frac{L-l+mL}{2t}}\frac{dk}{2\pi}s(k)l \\
\Delta C_m &= \int_{\frac{L-l+mL}{2t} \leq v < \frac{(m+1)L}{2t}}\frac{dk}{2\pi}s(k)((m+1)L-2tv(k)).
\end{aligned}
\end{align}

\begin{figure}[h!]
    \centering
    \includegraphics[width=0.5\textwidth]{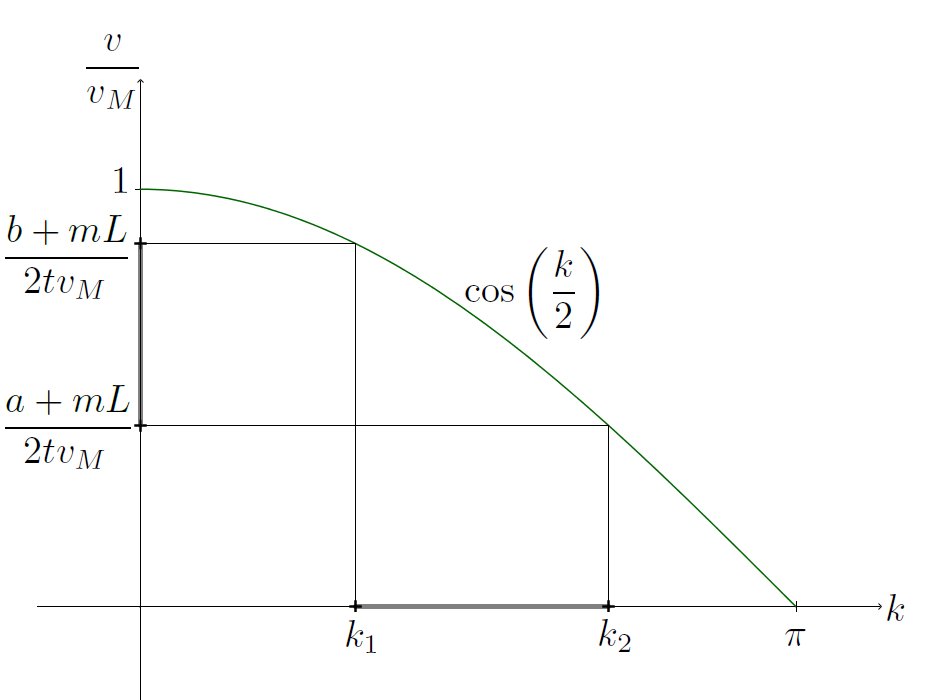}
    \caption{Integration boundaries $\frac{a+mL}{2t} \leq v_M\cos \left(\frac{k}{2}\right) < \frac{b+mL}{2t}$}\label{fig:Ising_integralBC}
\end{figure}

For the Ising chain N\'{e}el quench velocity (\ref{eq:Ising_v(k)}) these boundaries $\frac{a+mL}{2t} \leq v < \frac{b+mL}{2t}$ consist of one interval $[k_1,k_2]$, as illustrated in Figure \ref{fig:Ising_integralBC}, where $k_1=2\cos^{-1}\left(\frac{b+mL}{2t}\right)$ and $k_2=2\cos^{-1}\left(\frac{a+mL}{2t}\right)$, so
\begin{align}\label{eq:Ising_integralBC}
\int_{\frac{a+mL}{2t} \leq v_M\cos \left(\frac{k}{2}\right) < \frac{b+mL}{2t}}dk f(k) = \begin{cases}
0 & \mbox{if } t < \frac{a+mL}{2v_M} \\
\int_{0}^{2\cos^{-1}\left(\frac{a+mL}{2tv_M}\right)}dkf(k) & \mbox{if } \frac{a+mL}{2v_M}\leq t <\frac{b+mL}{2v_M} \\
\int_{2\cos^{-1}\left(\frac{b+mL}{2tv_M}\right)}^{2\cos^{-1}\left(\frac{a+mL}{2tv_M}\right)}dkf(k)  & \mbox{if } t\geq \frac{b+mL}{2v_M}. \end{cases} 
\end{align}
The integrands in (\ref{eq:EE_periodic_deltasum}) are all linear combinations of $s(k)$ and $s(k)v(k)$. The antiderivatives $I_s=\int dks(k)$ and $I_{vs}= \int dkv(k)s(k)$ of  (\ref{eq:Ising_s(k)}) and (\ref{eq:Ising_v(k)}), for the Ising chain N\'{e}el quench, are
\begin{align}
I_s(k) &= (2\ln 2 -1)k + 2i\left(\operatorname{Li}_2\left(-i\mathrm{e}^{-i\frac{k}{2}}\right)-\operatorname{Li}_2\left(-i\mathrm{e}^{i\frac{k}{2}}\right)\right) + \frac{i\pi}{2}k+\ln\left(\frac{1+\sin\left(\frac{k}{2}\right)}{1-\sin\left(\frac{k}{2}\right)}\right)\cos\left(\frac{k}{2}\right), \nonumber\\
I_{vs}(k) &= v_M\left(\frac{1-\sin\left(\frac{k}{2}\right)}{2}\right)^2\left(2\ln\left(\frac{1-\sin\left(\frac{k}{2}\right)}{2}\right)-1\right) - v_M\left(\frac{1+\sin\left(\frac{k}{2}\right)}{2}\right)^2\left(2\ln\left(\frac{1+\sin\left(\frac{k}{2}\right)}{2}\right)-1\right),
\end{align}
where $\operatorname{Li}_2(z)=-\int_0^z \frac{\ln(1-t)}{t}dt$ is the dilogarithm (Spence's function). These antiderivatives were chosen such that $I_s(0)=I_{vs}(0)=0$. (Despite the complex numbers in the expression, the imaginary part of $I_s(k)$ remains 0 for all $k\in[0,\pi]$). Denoting $I_s^o(x)=I_s(2\cos^{-1}(x))$ and $I_{vs}^o(x)=I_{vs}(2\cos^{-1}(x))$ the integrals in (\ref{eq:EE_periodic_deltasum}) become
\begin{align}
&\Delta A_m+\Delta B_m+\Delta C_m \nonumber\\
&=\begin{cases}
0 &\mbox{if }  t < mt_R\\
\frac{t}{\pi}I_{vs}^o\left(\frac{mt_R}{t}\right) - \frac{mL}{2\pi}I_s^o\left(\frac{mt_R}{t}\right) &\mbox{if } mt_r\leq t < mt_R+t_S\\
\begin{aligned} &\frac{t}{\pi}\left(I_{vs}^o\left(\frac{mt_R}{t}\right)-I_{vs}^o\left(\frac{mt_R+t_S}{t}\right)\right) - \frac{mL}{2\pi}I_s^o\left(\frac{mt_R}{t}\right) \\ &~~+ \frac{mL+l}{2\pi}I_s^o\left(\frac{mt_R+t_S}{t}\right)  \end{aligned} &\mbox{if } mt_R+t_S \leq t < (m+1)t_R-t_S \\
\begin{aligned} &\frac{t}{\pi}\left(I_{vs}^o\left(\frac{mt_R}{t}\right)-I_{vs}^o\left(\frac{mt_R+t_S}{t}\right) - I_{vs}^o\left(\frac{(m+1)t_R-t_S}{t}\right)\right) \\ &~~+ \frac{mL+l}{2\pi}I_s^o\left(\frac{mt_R+t_S}{t}\right)  + \frac{(m+1)L-l}{2\pi}I_s^o\left(\frac{(m+1)t_R-t_S}{t}\right) \\ &~~ - \frac{mL}{2\pi}I_s^o\left(\frac{mt_R}{t}\right) \end{aligned} &\mbox{if } (m+1)t_R-t_S \leq t < (m+1)t_R \\
\begin{aligned} &\frac{t}{\pi}\left(I_{vs}^o\left(\frac{mt_R}{t}\right)-I_{vs}^o\left(\frac{mt_R+t_S}{t}\right)  -I_{vs}^o\left(\frac{(m+1)t_R-t_S}{t}\right)\right)\\ &~~+ \frac{mL+l}{2\pi}I_s^o\left(\frac{mt_R+t_S}{t}\right)  + \frac{(m+1)L-l}{2\pi}I_s^o\left(\frac{(m+1)t_R-t_S}{t}\right) \\ &~~- \frac{(m+1)L}{2\pi}I_s^o\left(\frac{(m+1)t_R}{t}\right) -  \frac{mL}{2\pi}I_s^o\left(\frac{mt_R}{t}\right) +\frac{t}{\pi}I_{vs}^o\left(\frac{(m+1)t_R}{t}\right)  \end{aligned} &\mbox{if }  t \geq (m+1)t_R. \end{cases}
\end{align}
Inserting this into (\ref{eq:EE_periodic_deltasum}) then gives the entanglement entropy $S_l(t\in[nt_R,(n+1)t_R])$ for any $n\in\mathbb{N}$. 
\nocite{*}

\twocolumngrid

\bibliography{MPS}

\end{document}